\documentclass[sigconf]{acmart}
\usepackage[utf8]{inputenc}
\usepackage[english]{babel}
\usepackage[ruled,vlined,linesnumbered]{algorithm2e}
\usepackage{graphicx}
\usepackage{amsthm}
\usepackage{multirow}
\usepackage{enumitem}
\usepackage{tikz}
\usepackage{nomencl}
\usepackage{bbm}
\usepackage{verbatim}
\makenomenclature
\usepackage{tikz}
 
\usepackage[compact]{titlesec}
\titlespacing{\section}{0pt}{0.5ex}{1ex}
\titlespacing{\subsection}{0pt}{0.5ex}{0ex}
\titlespacing{\subsubsection}{0pt}{0.5ex}{0ex}

\copyrightyear{2022}
\acmYear{2022}
\setcopyright{acmlicensed}
\acmConference[WSDM '22] {Proceedings of the Fifteenth ACM International Conference on Web Search and Data Mining}{February 21--25, 2022}{Tempe, AZ, USA.}
\acmBooktitle{Proceedings of the Fifteenth ACM International Conference on Web Search and Data Mining (WSDM '22), February 21--25, 2022, Tempe, AZ, USA}
\acmPrice{15.00}
\acmISBN{978-1-4503-9132-0/22/02}
\acmDOI{10.1145/3488560.3498495}

\setlist[itemize]{noitemsep, topsep=1pt}

\usepackage[compact]{titlesec}
\titlespacing\section{0pt}{6pt plus 0pt minus 6pt}{0pt plus 0pt minus 1pt}
\titlespacing\subsection{0pt}{6pt plus 0pt minus 6pt}{0pt plus 0pt minus 1pt}
\long\def\comment#1{}
\long\def\comments#1{}

\begin{document}
\fancyhead{}

\setlength{\abovedisplayskip}{0pt}
\setlength{\belowdisplayskip}{0pt}


\title{Lightweight Composite Re-Ranking for Efficient Keyword Search with BERT}

\author{Yingrui Yang, Yifan Qiao, Jinjin Shao,  Xifeng Yan, Tao Yang}
\affiliation{%
  \institution{Department  of Computer Science, University of California at Santa Barbara, USA}
}
\email{ {yingruiyang, yifanqiao, jinjin\_shao, xyan, tyang}@cs.ucsb.edu}
\comments{  
}


\begin{abstract}
    

\comments{
Although considerable efforts have been devoted to transformer-based ranking models for 
document search, the relevance-efficiency tradeoff remains an important problem for ad-hoc ranking.
To overcome this challenge, }
Recently transformer-based ranking models have been shown to deliver high relevance 
for document search and the relevance-efficiency tradeoff becomes important for fast query response times.
This paper presents BECR (\textbf{BE}RT-based \textbf{C}omposite \textbf{R}e-Ranking), a lightweight
composite re-ranking scheme that combines deep contextual token interactions  and traditional lexical term-matching features. 
BECR conducts query decomposition and  composes  a query representation using  pre-computable 
token embeddings based on uni-grams and skip-n-grams, to
seek a tradeoff of inference efficiency and relevance. 
Thus it does not perform expensive transformer computations during online inference, 
and does not require the use of GPU.
This paper describes an evaluation  
of relevance and efficiency 
of BECR with several TREC datasets.


\end{abstract}



\maketitle

\comments{
\begin{table}
	\caption{Estimated Performance of Models on Keyword Based Search}
	\centering
	\begin{small}
		\begin{tabular}{c || c | c | c}
			\toprule
			Model 	& Target Query & Relevance & Inference Speed	\\
 			\midrule
			CONV-KNRM 	& keyword & + & +++\\
			BERT 	& Q\&A & ++ & + \\
			CEDR-KNRM & keyword & +++ & + \\
			PreTTR & keyword & ++ & ++ \\
			TK & Q\&A & ++ & ++ \\
			ColBERT & Q\&A & + & +++\\
			BECR & keyword & +++ & +++\\
			\bottomrule
		\end{tabular}
		\label{tab:comparison}
	\end{small}
\end{table}
}

\section{Introduction}
\label{sec:intro}
The advent of transformer architectures (e.g., BERT) and pretraining techniques rejuvenated neural ranking studies using deep contextual models (e.g. ~\cite{ cedr, bertmaxp,  Lin2020PretrainedTF}), leading to a higher relevance score in text document search. 
However, the cost of runtime transformer-based inference is still very expensive.
To address this limitation, recent studies examined the efficiency of transformer-based models during model 
inference~\cite{prettr, bertmaxp, tk, colbert}. 
Nonetheless, devising a fast online ranking scheme with a relevance comparable to those of expensive transformer-based 
schemes remains a 
critical problem, especially
on a CPU-only server without GPU support.
\comments{
Previous research on transformer-based neural ranking models usually experiment with keyword queries or Q\&A queries 
that require reading-comprehension such as TREC-CAR dataset~\cite{treccar}.
It is intuitive that a fine-tuned transformer model would perform well through deep semantic understanding of a query-document pair jointly. 
Unfortunately, such query-time semantic understanding is highly time consuming and thus only applicable to a relatively small dataset. 
For a large dataset, such time-consuming transformer-based computation is deemed impractical~\cite{Lin2020PretrainedTF}.
}
To reduce the complexity involved in the transformer computation, simplified BERT based~\cite{bert} architecture was previously proposed to reduce the burden of jointly encoding query-document pairs, including the dual encoder based rankers that separately computes the query and document 
representations~\cite{colbert,dcbert}. 
With dual encoder models, the BERT representations of documents become independent of the queries and can thus be pre-computed. 
However, expensive transformer computation still needs to be conducted during online inference 
as queries are dynamic. 
\comments{
Furthermore, despite the improved efficiency, these aforementioned models are still not able to achieve an NDCG~\cite{NDCG} relevance competitive to 
that 
of CEDR~\cite{cedr} on ad-hoc ranking tasks.

This paper considers the efficiency optimization for short  ad-hoc search query, 
motivated by the fact that 
To overcome these obstacles, on top of the previously proposed dual-encoder design~\cite{prettr,colbert}, 
}

Considering that
short keyword queries account for a majority of web search traffic~\cite{2017GoogleTraffic}, where the average number of words in user queries of popular search engines
is between two and  three~\cite{JS06},
our approach takes a further step on top of 
the previously proposed dual-encoder design~\cite{prettr,colbert}. 
We propose a lightweight neural re-ranking scheme named BECR  that 
composes a query representation using pre-computable token embeddings based on uni-grams and skip-n-grams.
This representation composition does not require expensive transformer-based inference for an input query during runtime,
and thus GPU support is not necessary.
In addition, 
BECR leverages the classical lexical matching features to offset the relevance degradation from the above approximation,
and  use hashing~\cite{JiWWW2019} and distilling~\cite{prettr} to reduce storage cost. 

Our evaluation with several TREC datasets shows that this re-ranking scheme
can deliver decent  relevance performance compared to the other baselines.
Because BECR's online computation complexity is significantly smaller, 
the main contribution of this work is 
to provide  a lightweight and CPU-friendly online neural inference scheme to strike an efficiency and relevance tradeoff. 
Since adding a high-end GPU incurs significant  monetary cost and energy consumption,
the design of BECR can be attractive for fast query processing on servers without GPU support.  

\section{ Background and Related Work}
\label{sec:back}

The problem of top $K$ text document re-ranking is defined as follows:
given a query $q$ and candidate documents,  
form text features for each document and rank documents based on these features.
We focus on ad-hoc search where a query contains several keywords.
\comments{
An example of such applications is web search, where the average number of words in user queries of popular search engines 
is two or three~\cite{SMH99,JS06}. The proposed techniques are not targeted at natural language queries which are typically long.
}

Large-scale search systems typically deploy a multi-stage ranking scheme.
A common first-stage retrieval model uses a simple algorithm such as BM25~\cite{Jones2000} to collect top documents. 
The second or later stages use a more advanced algorithm to further rank top documents.
This paper focuses on re-ranking in the second stage for a limited number of top documents.
Below we review the previous studies in this field and highlight the key difference of our method. 

\textbf{Deep contextual ranking models}.
Neural interaction based models such as
DRMM~\cite{drmm} and (Conv-)KNRM~\cite{convknrm,knrm} have been proposed to leverage word embeddings for 
ad-hoc search. Compared to traditional ranking methods such as LambdaMART~\cite{Liu2010LearningTR}, these models provide a good NDCG~\cite{NDCG} score on TREC datasets.
However, their contextual capacity is limited by using non-contextual uni-gram embeddings. 
In recent years, the transformer based model BERT has been adopted~\cite{bert} to re-ranking tasks~\cite{Nogueira2019PassageRW,birch,bertmaxp, monobert, parade, Hofsttter2021IntraDocumentCL}. The merger of BERT models with interaction based neural IR models has also been investigated~\cite{cedr, tk, ck}. 
Among them, several models~\cite{cedr,parade,bertmaxp,birch} have achieved impressive relevance on keyword based queries. Work exploring zero-shot learning or weak supervision also demonstrates competitive performance on such ad-hoc datasets 
(ClueWeb and TREC Disk 4\&5)~\cite{monot5,Sun2021FewShotTR, Zhang2020SelectiveWS, Zhang2020ALB, Yilmaz2019CrossDomainMO}.

\textbf{Efficiency optimization for  transformer-based rankers}. 
Although the aforementioned transformer based  ranking achieves remarkable relevance, such an architecture
is  extremely expensive for online search systems due to 
the strict latency requirement~\cite{Lin2020PretrainedTF}. To this end, the dual-encoder or poly-encoder design that separately encodes the query and documents has been 
studied~\cite{Zhan2020LearningTR,Chen2020DiPairFA, colbert, Reimers2019SentenceBERTSE, Karpukhin2020DensePR,dcbert,Humeau2020PolyencodersAA, 
Zhan2020RepBERT, twinbert}. Among them, 
ColBERT~\cite{colbert} applies a BERT based dual-encoder design on re-ranking and retrieval. 
MORES~\cite{Gao2020ModularizedTR} framework separating the representation and interaction phase of query and documents achieves 120x speed up in inference compared to a BERT ranker.
TwinBERT~\cite{twinbert} is a lightweight transformer based dual-encoder distilled from a BERT cross-encoder teacher model. 
DC-BERT~\cite{dcbert} applies a dual-encoder BERT on document retrieval for open domain Q\&A tasks. 
These rankers greatly shorten latency through offline pre-computation of document representations 
while performing transformer computation to generate query representation at runtime.

In this paper, we adopt a dual encoder framework while further simplifying the online inference by decomposing 
queries to leverage pre-computed uni-gram and skip-n-gram embeddings.
A skip-n-gram~\cite{guthrie2006-skipgram} is a set of words that appear in a document within certain appearance proximity constraint.
To offset the relevance loss due to the above approximation, 
BECR  combines semantic matching with lexical text matching signals such as  
term-frequency based features from BM25, and word distance based proximity~\cite{Bai08,Sigir14Zhao}.
Thus it is also motivated by the work~\cite{Mitra2017LearningTM,Shao2019PrivacyawareDR,Gao2020ComplementingLR, Luan2020SparseDA} that 
investigates non-transformer based neural ranking with traditional text features. 

There are other lines of work that improves the efficiency of transformer architectures, including
compact attention based transformer architectures~\cite{tk,ck,prettr}, early exiting~\cite{Xin2020SUSTAINLP}, 
cascading~\cite{Hofsttter2021IntraDocumentCL, Soldaini2020TheCT}, quantization~\cite{2019QuantizedBERT},
and sparse transformers ~\cite{Jiang2020LongDR, tkl}. As they focus on the transformer model architectures, our proposed techniques are orthogonal to these models.
Our work has also leveraged embedding approximation based on locality-sensitive hashing~\cite{JiWWW2019} to
improve efficiency.
Other related studies on efficiency and relevance trade-offs for transformer-based ranking includes models that learns the importance or likelihood of tokens in query and documents~\cite{deepCT, tilde, epic}.
In the design of BECR's ranking formula, we are inspired by a recent learning-to-rank study~\cite{Zhuang2020InterpretableLW}
that advocates additive models for better ranking interpretability. 
This additive nature simplifies  the  incorporation of other application-specific ranking features. 

\section{The Proposed Framework}
\label{sec:method}

\begin{figure*}
  \includegraphics[width=1\textwidth]{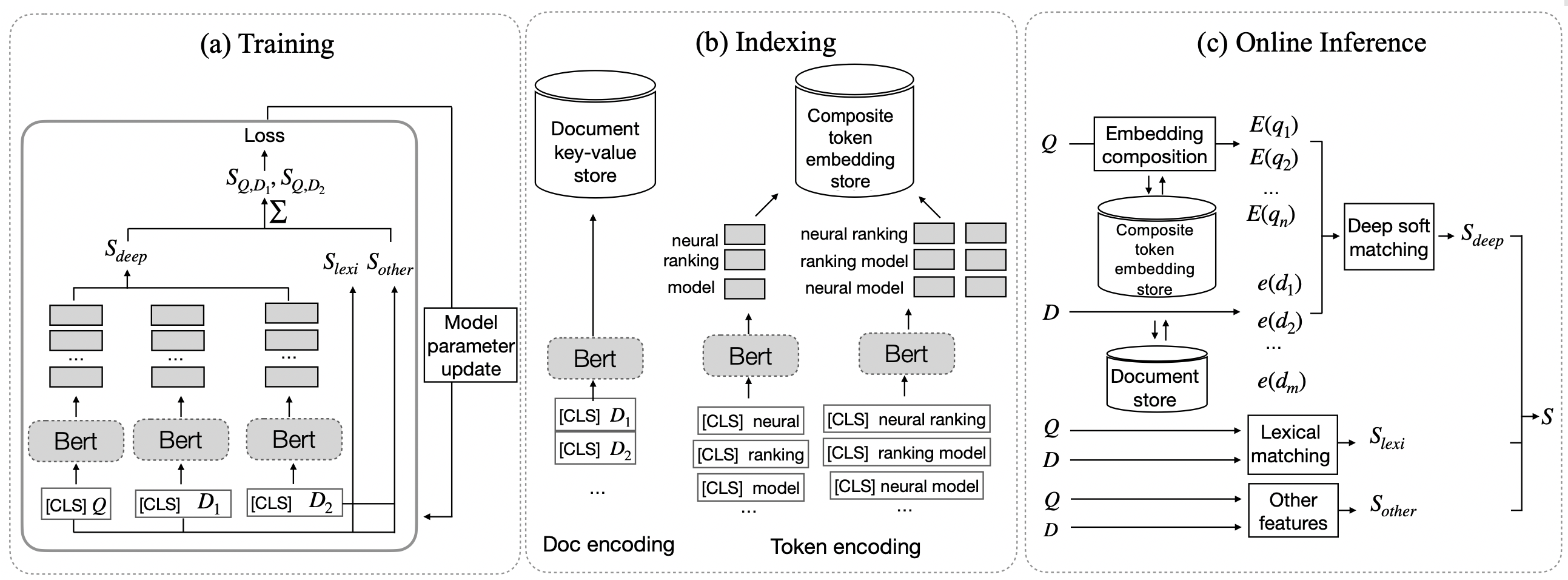}
  \caption{Training, Offline Processing and Online Inference in BECR}
  \label{fig:BECR}
\end{figure*}

This section presents our design considerations and the BECR re-ranking framework.
Figure~\ref{fig:BECR} illustrates the ranking structure of BECR with tri-encoding of documents, queries, and tokens.
BECR separately encodes a query and a document, then considers the term-level 
interaction to capture semantic relatedness of a document with a query, and combines that with other types of relevance signals. 

\subsection{Embedding Composition for Queries} 
\label{sec:method-tbe}

As discussed, previous dual-encoder rankers decouple the query and document encoding, and encode a query using BERT during online inference.
During the training time, a BERT query encoding model is trained.
During the indexing time, a query encoding model is stored with learned parameters.
During the inference time, a ranker feeds a query through the pre-stored BERT query encoding model to produce a set of contextual embeddings for query terms.

To improve the online efficiency, we propose to approximate the deep representation of a query through a composition 
of pre-computed token embeddings based on uni-grams and skip-n-grams. 
Given a query composed of a set of terms  $Q= \{ q_1, q_2,\cdots, q_n\}$, and a document $D$ with $m$ terms, $d_1, d_2,...,d_m$,
we decompose a query into a set of semantic token groups and denote it as $T$. 
In our evaluation, $T$ is a set of uni-gram tokens and token pairs within a context window size such as  $|window|=3$ to control the number of pairs, 
meaning we only consider the token 
pairs within distance 3 inside the query.

The contextual embedding of each query term  $q_i$ in a given query is defined as 
\comments{
In our evaluation of English document search, we let term $q_i$  be  a  word.
We decompose a query into a semantic token/token pair
set called $T$ that contains uni-grams and pairs of tokens from the query. In our evaluation, we use token pairs with the default context window size such as  $|window|=3$, meaning we only consider the token pairs within distance 3 inside the query.
}

\begin{align}
     E(q_i) = \sum_{t \in T \land contain(t, q_i) }  \frac{ w_{t} }{
 \|w\|_1}
 e(q_i^t)
 \label{eq:tok_encoding}
\end{align}
where $e(q_i^t)$ represents the pre-computed embedding vector of 
term $q_i$ in token group $t$,
and $w_t$ is the weight factor for  $t$.
Predicate $contain(t, q_i) $ means that  token group $t$ semantically relates to term $q_i$.

In our evaluation described in Section~\ref{sec:eval}
where a query term is a word token, and $t$ is either a word based token or a pair of such tokens,
we simply define $contain(t, q_i)$ to be true when $t$ contains $q_i$.
$\|w\|_1$ is the 1-norm of all weights $w_t$ used so that they are normalized as a weight distribution.
\comments{
\begin{figure}[htbp]
    \centering
   \includegraphics[width=0.4\textwidth]{../img/q_weight_0130.pdf}
    \caption{BECR: Composite Re-ranking with BERT}
    \label{fig:q_encode}
\end{figure}
}
A uni-gram is weighted smaller than a token pair, while we impose a scaling factor for each token pair
based on the word distance in a query. Specifically  the weight of
token pair $(a, b)$ is $\frac {1}  {span(a,b)}$ where $span(a,b)$ is the word distance of words $a$ and $b$. The weight for the uni-gram will be defined as  $\frac {1}  {|window|+1}$.

For example, given query ``\textit{neural ranking model}'', the token group set $T$  for this query is 
$\{$neural, ranking, model, neural-ranking, neural-model, ranking-model$\}$.
Assume all these token pairs and uni-grams have the embeddings available from the index.
The embedding vector of term ``neural'' in  this query will be: 
\begin{align*}
    &( w_{1}\cdot e(\mbox{neural}^{\mbox{neural}}) + w_{2}\cdot e(\mbox{neural}^{\mbox{neural-model}}) \\
    & + w_{3}\cdot e(\mbox{neural}^{\mbox{neural-ranking}}) )\frac{1}{\|w\|_1}
\end{align*}
\noindent
where $w_1$=0.25, $w_2$=0.5, $w_3$=1.
Expression $e(\mbox{neural}^{\mbox{neural-model}})$ is the embedding of query term ``neural'' under token group
		``neural-model'' computed during the offline indexing time.

Our current evaluation uses heuristically-determined weights for the above composition and we leave the weights exploration through machine learning to future work.

Figure~\ref{fig:BECR} illustrates training, offline indexing, and online inference with query embedding composition in BECR using an example. 

\comments{
To understand the effect of the weight factors on the model performance, in the Evaluation section, we explore extreme cases where either uni-gram or bigram weights are pushed to 0. We found that the token weights are not a determining factor of the model's performance. Hence in this work, we stick to this heuristically determined weights.

An example of the three stage process incorporating the decomposition and composition mechanism is illustrated in Figure~\ref{fig:BECR}. We train a BECR model from end-to-end, and derive token encodings using the fine-tuned BERT parameters in BECR during indexing, then approximate the query representations during inference using pre-computed token/token pair encodings.
}

During the training time as depicted in Fig.~\ref{fig:BECR}(a), 
we input a set of training queries with their matched documents to the BECR model to learn model parameters in an end-to-end manner. 
Given training query $Q$ and matched documents $D_1$ and $D_2$, assuming $D_1$ is more relevant than $D_2$ for $Q$, the goal of training 
is to minimize a pairwise softmax loss. There are two methods to train the model. 
1)
We don't decompose queries, and  input the whole query and document separately into BECR using $[CLS] Q$ directly. 
2) We decompose the query and compose the query embeddings based on Equation (\ref{eq:tok_encoding}). 
In our evaluation, 1) is more effective for ClueWeb09-Cat-B and Robust04 with limited number of training queries. 2) is used for MS MARCO with more training queries.
 
During the indexing time as depicted in Fig.~\ref{fig:BECR}(b), 
we select a set of desired token groups (e.g. uni-gram tokens and  token-pairs),  and feed them
to the trained model to pre-compute their embeddings. The results (after compression) are saved in a key-value store called
{\em composite token embedding store} for online access.

During the inference time as depicted in Fig.~\ref{fig:BECR}(c), we decompose
 a query into a set of tokens,  represent a query term embedding using token groups (e.g. uni-gram tokens, or token pairs), 
fetch the embedding of the token groups from the composite token embedding store.
If some token pair embeddings are not found from the composite token embedding store, we can ignore them because we have at least the uni-gram embeddings to work with. 


\subsection{Online Composite Re-Ranking}
\label{sec:method-crl}

We divide ranking signals into 3 components: deep contextual soft matching, lexical matching, and other features. To derive the final ranking score, we opt to use an additive combination of these components for simplicity. To get the component score, we use simple feed forward layer to combine the features. Note that others can easily add more components based on their applications or use a more complex architecture for composition.

To derive these weight parameters used in each component,  the overall framework along with its component weights is trained end-to-end jointly. The final ranking score $S$ is the addition of a summary score from each component: 
\begin{align*}
    S = S_{deep} + S_{lexi} +  S_{others}.
\end{align*}

We use symbols $\alpha$, $\beta$ and $\gamma$ to represent the weight parameters in the three components, respectively.

\textbf{The deep soft matching component}.
The deep semantic matching component adopts a deep learning architecture similar to the CEDR-KNRM model~\cite{cedr,knrm}. 
Compared to CEDR-KNRM, we modify the final scoring such that it is decomposed into a set of subscores where each of them corresponds to a query term.  


Note that a BERT-based embedding contains $L$ layers,
and  we denote the representation of query term $q_i$ under layer  $l$ as $\boldsymbol{E}_{l,q_i}$. 
The similarity score $c_{l,i,j}$ between query term $q_i$ and document term $d_j$ under layer  $l$ is computed as the cosine similarity between 
$\boldsymbol{E}_{l,q_i}$ and $\boldsymbol{E}_{l,d_j}$. 
\begin{align*}
    c_{l, i,j} = & cosine(\boldsymbol{E}_{l,q_i}, \boldsymbol{E}_{l,d_j}),
\end{align*}

For each layer $l$, we group the similarity subscores for each query term using the $K$ RBF kernels following \cite{knrm}, 
where $\mu_k$, $\sigma_k$ are the parameters of the k-th kernel. The subscore ($s_{l, i, k}$) for the query term $q_i$ under layer $l$ and  kernel $k$ is defined as:
\begin{align*}
     s_{l,i, k} = & \log \sum_{1\leq j \leq m} (-\frac{(c_{l,i,j }-\mu_k)^2}{2\sigma_k^2}).
\end{align*}
where this document contains $m$ term. Note that in our model, $\mu_k$, $\sigma_k$ are trainable hyper-parameters hence there is no exact matching kernel as specified in KNRM model. We rely on the lexical component to capture exact matching information.

For each query term $q_i$, there are  a total of $L*K$ features given $L$ BERT layers and $K$ kernels. The soft semantic matching  score is the sum of all term subscores.
\begin{align*}
    S_{deep} = & \sum_{1\leq i \leq n} S_{deep, i} \\
    S_{deep, i} = & \sum_{1\leq k \leq K, 1\leq l \leq L} \alpha_{k,i,l} s_{l,i, k}
    \label{sdeep}
\end{align*}
where $\alpha_{k,i,l}$ is a weight parameter for subscore   $s_{l,i, k}$.


 

\textbf{The lexical matching component}.
We include a set of lexical matching features to complement the deep contextual soft matching component. 
In our evaluation, we use BM25~\cite{Jones2000} for query words that appear in the title and body separately,
and  BM25 for query word pairs that appear in the title and body, plus other query word pair based
proximity features~\cite{Bai08,Sigir14Zhao,Shao2019PrivacyawareDR} with the minimum, maximum,
and average of the squared min distance reciprocal of query word
pairs in the title or in the body. The lexical score $S_{lexi}$ is obtained through a linear combination of all these signals.

\comments{
\begin{align*}
    S_{lexi} = & \sum_i \beta_{i} f_{i} 
\end{align*}
where  $f_{i}$ is the $i$-th lexical matching feature and $\beta_{i}$ is its weight parameter. 
}

\textbf{Other features}. 
We add the last layer's [CLS] token representation of each document as document-specific features. 
There are  other query independent or dependent features for each document used in traditional text ranking and web search applications.
The examples of such signals include document quality,  freshness and document-query  click through rate.
For the ClueWeb09-Cat-B dataset used in our evaluation, we have used a page rank score for each document.
The other ranking component $S_{others}$ is the linear combination of all these features.
The online inference can obtain these features from  the inverted index or from a separate key-value store.
\comments{
\begin{align*}
    S_{others} = & \sum_i \gamma_{i} \cdot z_i
\end{align*}
where $z_i$ is the $i$-th other features to be added,
and $\gamma_{i}$ is its weight parameter. Note that all the bias terms in the formula are omitted to increase readability.}

Section~\ref{sec:expl} in the appendix gives an example of ranking two documents for a query  
with BECR and illustrates the above scoring components and the trained  weights.

\subsection{Storage Cost and Reduction Strategies}
\label{sec:storage}

As pointed out in~\cite{Lin2020PretrainedTF}, 
efficient neural ranking models need to take into account the 
storage cost 
in addition to query latency and effectiveness.
The contextual nature of the transformer architecture implies that the same word in documents can have different contextual representations. Thus there is a significant
storage requirement for hosting embedding vectors for a large number of documents. 


\begin{table}[htbp]
    \caption{Sample online embedding storage cost for BECR}
    \centering
    \begin{small}
    \begin{tabular}{|c|c|c|}
        \hline
                &Doc embeddings & Token embeddings   \\
        \hline
 Original space & $4(mL$+$1)768D $ &  $4L(V$+$2H)768$  \\
        \hline
 Compressed space & $(mL'\frac{b}{8}+4*768)D$ &  $L'(V+2H)\frac{b}{8}$  \\
         \hline
$m$=$857$,$V$=$14.5$M,$H$=$467$M & Orig: 1,711TB  &   Orig: 37.9TB \\
$L'$=$5$, $b$=$256$, $D$=$50$M, $L$=$13$ &Compress: 7.0TB  &  Compress: 152GB  \\
         \hline
    \end{tabular}
    \end{small}
    \label{tab:storagecost}
\end{table}

{\bf Storage cost for document embeddings.}
Assuming the set of $D$ documents is not updated frequently, we encode each document using BERT with $L$ layers and save it in a key-value store. 
Table~\ref{tab:storagecost} gives an estimation of storage size in bytes for hosting embeddings of documents and semantic tokens in BECR with an example (before and after compression with an approximation).
We assume the BERT-base model is used with the vector dimension of 768.
To deal with long documents with more than 512 tokens, following \cite{cedr}, we split the documents into even length texts and learn the token-level representation piece by piece. The [CLS] representation of the whole document is calculated by averaging the [CLS] representations of all text pieces from the document.
Denoting the average length of a document as $m$ tokens for web documents, 
the storage space plus one extra [CLS] embedding vector is $4*(mL+1)*768$ bytes per document, as the values in a BERT term embedding vector is a 4-byte floating point.
For a database with 50M documents and average length $m=857$, following the characteristic of ClueWeb09-Cat-B without spam document removal, 
the storage per document is 34.2MB  and the total storage space is about 1,711TB. 
That is huge and unrealistic, even the size of this dataset can be further reduced by 
removing  some spam documents as we did in Section~\ref{sec:eval}.

The work of Ji et al.~\cite{JiWWW2019} proposes  an embedding  approximation  based on locality-sensitive hashing (LSH).
PreTTR~\cite{prettr} uses term representation compression, 16-bit floating point  numbers, and distilling with a limited number of BERT layers to 
reduce the storage demand.  ColBERT~\cite{colbert} uses vector dimension reduction to decrease the embedding vector size  from 768 to 128.
BECR adopts some of these ideas.
Our evaluation shows in Section~\ref{sec:eval} that decreasing the number of layer $L$ from 13 to 5 with a proper selection of layers
does not lead to a noticeable impact on NDCG@5 relevance while there is  a 0.01 point loss in NDCG@10 and NDCG@3.
Thus choosing 5 layers can be a feasible strategy.

\comments{
BECR  adopts some of these ideas to use 5 output layers in our evaluation instead of standard 13 layers for the BERT base model
and uses an approximation  method based on locality-sensitive hashing~\cite{lsh}. 
}

The major space reduction comes from LSH approximation.
We reduce the size of each embedding vector from 768$\times$4 bytes to only $b$ bits while preserving the accuracy of cosine similarity between two approximated vectors. 
The estimation of cosine similarity between two embedding vectors uses the Hamming distance between their LSH footprints. 
\comments{
The vector approximation procedure with LSH
is described as follows. 
\begin{itemize}
    \item Step 1. To compute a $b$-bit LSH footprint, we independently sample $b$ random value $r_i$ following a multivariate normal distribution from the vector space of the embeddings.
    \item Step 2. For each document token embedding vector $\it{v}$, we generate b-bit footprint by 
    \begin{align*}
        LSH(\it{v}[i]) = \mathbbm{1}\{r_i\it{v} > 0\}
    \end{align*}
    \item Step 3. After repeating Step 1 for all the document and query tokens, we use Hamming distance between their LSH footprints, to estimate the cosine similarity between tokens.
\end{itemize}
}
In Table~\ref{tab:storage} of Section~\ref{sec:eval:design},
we experiment with the tradeoff between NDCG scores and footprint dimension $b$. 
Our evaluation shows that the LSH footprint dimension with 256 bits (namely 32 bytes) can deliver a highly competitive relevance. 

Last row of Table~\ref{tab:storagecost} lists
the document embedding storage space after LSH hashing and setting $L'=5$.
The storage space need is reduced to about 138.7KB per document based on expression  
$857\times5\times256/8 + 4\times768$ and the total is  about 7.0TB for a 50 million document set with average length 857.
Considering Samsung and Seagate have produced an internal 2.5'' SSD drive with a large capacity varying from 8TB to 32TB for servers,
and each server can host multiple internal drives, the above storage cost is reasonable for a production environment. 

\comments{
\textbf{Further reduction with less layers}. As document storage is linear with respect to the number of representations a model use. 
Currently our model uses 
the embeddings from all 13 layers in BERT-base model. It is feasible to decrease the number of representation to 5 
(which leads to 6.85TB document storage in total) as shown in Section ~\ref{sec:eval} without a noticeable impact on performance.   

}
\comments{
One of the key contributions of our framework is the flexibility to pre-compute all the BERT related representations offline. Since queries are dynamic, 
in order to learn the query representations offline, we can use BERT to perform token-level encoding on the WordPiece token~\cite{wordpiece} offline and 
generate query representations from token-level presentations in the online system. 
}

{ \bf Storage cost for the composite token embedding store.}
This key-value store can be accessed by token group identifiers. In our evaluation, this store  saves embeddings for
uni-gram tokens and  token pairs. 
Table~\ref{tab:storagecost} gives an estimation of storage size in bytes when the number of uni-grams tokens and token pairs is $V$ and $H$ respectively.
Notice that each token pair has two vectors to represent its embedding  while an uni-gram token has one vector.
Without LSH approximation, the total number of bytes needed for token embedding storage would be  $4L(V+2H)\times768$ with a 32-bit floating point number.

We use the characteristics of the ClueWeb09-Cat-B dataset to show the cost of token embedding storage.
This dataset has about  V=14.5M  uni-grams that appear more than once in documents, and H=467M word-pairs that appear more than once and within window size 2.
The total cost is 37.9TB and
thus aggressive compression with an approximation  is desired.
By using the 256-bit LSH footprint and 5 BERT layers for each token, the above storage is reduced to $32L'(V+2H)$ bytes, which is about 152GB.
If we include all uni-grams and word pairs within window 2, then $V=32.4$M and $H=940.3$M. The space cost is 305.5GB, which is still affordable.

Note that there are many ways to further adjust the choices of uni-grams and skip-n-grams based on an application dataset and its relevance study results for training instances.
For example, one could add some restriction on the token set  based on the BERT dictionary which  
uses WordPiece~\cite{wordpiece} tokenizer and  the number of BERT uni-grams is around 32,000.

\subsection{Discussion on Online Inference}
\label{sect:discussonline}


Since  a query term is represented by a set of token groups (e.g. tokens and token pairs) and
the similarity angle of two LSH-based bit vectors is proportional to their Hamming distance scaled by the total number of bits. 
We approximate  the cosine similarity between a query term and a document term in the LSH setting
using the weighted linear combination of the cosine similarity between each of  these token groups and a document term.
These weights follow the ones used in Equation ~\ref{eq:tok_encoding}.

When processing each online query, we can break the inference into two steps. 
First, a serving system looks up the LSH footprint of the query token groups and  the matched documents separately. Afterwards, it
derives soft interaction between each selected token group and document terms
through pair-wise hamming distance computation.
Second, given the soft interaction scores of document term embeddings  and query token embeddings based on their
LSH footprints, we use the weighted linear combination  to derive the similarity scores between query terms and document terms.
Separately, the lexical features and other features would be passed down from the document retrieval stage of the search system. 
Then given the contextual score, lexical features, and other feature subscores, we compose the final ranking score.

We have implemented a key-value store to host document and composite token embeddings, and in our experiment with
a Samsung 960 NVMe SSD, fetching one document embedding vector randomly costs 
0.17 milliseconds (ms) with 5 layers, and 0.21ms with 13 layers, while fetching a token embedding takes about 0.058ms in a sequential I/O mode.
Thus fetching 150 different document embeddings  and about ten or less query token embeddings takes about 26ms in total for the 5-layer setting and 32ms for 13 layers. Such a cost
is reasonable to accomplish interactive query performance. 

\comments{
Mayuresh's number based on Yifan's computer

0.15 milliseconds while fetching a token embedding takes about 0.05 milliseconds.
Thus fetching 150 different document embeddings  and a few token embeddings takes about 23 milliseconds in total with a sequential I/O mode.

Jinjin's number

For the access time in my ssd (Samsung 970 EVO V-NAND 1TB),
4KB is 0.058ms/0.036ms/0.021ms with maximum queue size being 1/2/4.
130KB is 0.17ms/0.15ms/0.11ms with max queue size being 1/2/4.
300KB is 0.21ms/0.20ms/0.17ms with max queue size being 1/24.
If we use 0.058ms, 0.17ms, and 0.21ms here, that assumes  one request a time/sequential I/o in the worst case
 (namely because there are many concurrent queries in a real web site,  we cannot use parallel access for one query, otherwise other queries would have no resource).
Jinjin's number based on his Samsung disk
}

The dominating part of online inference time complexity is the computation of query-document term interaction and kernel
calculation with $L$ BERT layers, which is $\Theta(n \cdot m \cdot p + n \cdot m \cdot K \cdot L)$
where $n$ is the number of terms per query,  $m$ is the average number of terms per document, $p$ is the size of the embedding vector for each term,
$K$ is the number of kernels, and $L$ is the number of BERT layers.  With $b$-bit LSH approximation, 
the above cost can be reduced to $\Theta(n \cdot m \cdot f(b) + n \cdot b \cdot K)$ where $f(b)$ is the cost to count the number of non-zero bits
in computing the hamming distance of two binary LSH footprints. Together with an efficient implementation of non-zero bit counting, 
the above complexity reduction shortens  online inference time of BECR significantly.

\comments{
The online inference of BECR does not involve transformer computation.
Since adding a high-end GPU incurs significant  monetary cost and energy consumption, 
many servers do not have GPUs and the time efficiency of BECR online inference can be attractive for search systems running on such servers. 
}

\section{Evaluations}
\label{sec:eval}

Our evaluation addresses the following research questions.
\begin{itemize}[leftmargin=*]
\item{\textbf{RQ1}}: How effective is BECR compared to the recent state-of-the-art  algorithms in terms of relevance?

\item{\textbf{RQ2}}: What is the inference speed during query processing and what is the computation cost of BECR compared to the baselines? 
\item{\textbf{RQ3}}: How effective are different design options related to the proposed query decomposition and embedding generation?   
\item{\textbf{RQ4}}: How does each component of BECR additive scoring contribute to its overall relevance?
\item{\textbf{RQ5}}: What is the storage cost of BECR with embedding compression through LSH approximation?
How does LSH mapping  and a smaller number of BERT layers used affect the relevance?
\end{itemize}

\comments{
\vspace*{-3mm}
\begin{table}[htbp]
    \caption{Summary of Evaluation Data}
    \centering
    \resizebox{1.02\columnwidth}{!}{
    \begin{small}
    \begin{tabular}{|c|c|c|c|c|c|}
        \hline
   		Dataset & Domain & \# Query & \# Doc & Query Length & Mean Doc Length \\
        \hline
        ClueWeb09 & Web & 149 & 50M & 1-5 & 857 \\
        Robust04 & News & 250 & 0.5M & 1-4 & 479 \\
        MS MARCO Dev & Q\&A, passage & 6980 & 8.8M & 2-15 & 57\\
        TREC DL19 & Q\&A, passage & 43 & -- & 2-15 & -- \\
        TREC DL20 & Q\&A, passage & 54 & -- & 2-15 & --\\
        \hline
    \end{tabular}
    \end{small}
    }
    \label{tab:data}
\end{table}
\vspace*{-3mm}
}

\subsection{Setting}
\label{sec:inference:data}
\noindent

\noindent
\textbf{Evaluation setup}.
BECR is implemented and evaluated in Python PyTorch\footnote{Our code is available at \url{https://github.com/yingrui-yang/BECR}}. We also implement a C++ version to prototype its inference efficiency without GPU for online ranking. 
The three standand TREC datasets for ad-hoc search are used for evaluation and they are Robust04, ClueWeb09-Cat-B, and MS MARCO
as described in Appendix~\ref{sec:data}.
\comments{
Following previous work~\cite{cedr}, for Robust04 and ClueWeb09, we re-rank the 150 documents per query retrieved by Indri initial 
ranking in the Lemur system\footnote{http://www.lemurproject.org/}. For MS MARCO, we re-rank top 1000 documents per query retrieved by BM25.
}

We will report the commonly used  relevance metrics for each dataset, including NDCG@k, P@k and MRR. 
To compare two models, we perform paired t-test comparing average query-level metrics and marked the results with statistical significance at confidence level 0.95. 
To measure the time cost, we report the document re-ranking  time and the number of performed floating point operations (FLOPs).
We exclude query processing time spent for index matching, and initial stages of ranking. 
We list the main storage cost in bytes for hosting document and token embeddings.
The GPU server used is the AWS ec2 p3.2xlarge instance with 
Intel Xeon E5-2686v4 2.7 GHz,
61GB main memory and a 
Tesla V100 GPU with 16GB GPU memory. 
The CPU server used 
is Intel Core i5-8259U with up-to 3.8GHz, 32GB DDR4, and 1TB  NVMe SSD. 

\noindent
\textbf{Training setup}.
All models are trained using 
Adam optimizer~\cite{Kingma2015AdamAM},
and pairwise softmax loss over per-query relevant and irrelevant document pairs.
Positive and negative 
training documents are randomly sampled from query relevance judgments. 
For Robust04 and ClueWeb09-Cat-B, we use 5-fold cross-validation and each model runs up to 100 epochs, with 512 training pairs per epoch. The batch size ranges from 1 to 8 depending on the model used due to the limit of GPU memory. We use gradient accumulation and perform propagation every 16 training pairs. The learning rate for non-BERT layers is 0.001, and 
$2\times 10^{-5}$ for BERT layers.
For MS MARCO passage reranking, we trained the model on training pairs for 200,000 iterations with batch size 32, 
learning rate $3\times 10^{-6}$.
We evaluated our model on MS MARCO dev set, TREC DL19 and TREC DL20 datasets.



As the number of queries in the training set is limited for Robust04 and ClueWeb09-Cat-B, we fine-tune the models in a step-wise approach. First, we trained a vanilla BERT ranker. Then we load the weights from BERT as a starting point to train our BECR model. We also noticed that for Robust04 and ClueWeb09-Cat-B, training a model without query decomposition and evaluate with query decomposition during inference is less prone to over-fitting. On MS MARCO as the training samples are abundant, we directly trained our model with query decomposition. Note that all of the fine-tuning processes are end-to-end, i.e, we unfreeze the layer parameters in all of the fine-tuning steps.

{\bf Zero-shot learning.} With the model trained on MS MARCO, we performed zero-shot evaluation on ClueWeb09-Cat-B and Robust04, and
observed a degradation of around 5 percentage points in NDCG@20 compared to the model trained on the two dataset respectively.
Thus this paper has  not adopted the zero-shot training strategy and our future work is to investigate this further. 

\noindent
\textbf{Baseline setup}. The baselines 
include CONV-KNRM~\cite{convknrm}, (vanilla) BERT ~\cite{bert},  and CEDR-KNRM~\cite{cedr}. ColBERT~\cite{colbert} is also compared since it is a model focusing on efficiency as well. We have leveraged the CEDR~\cite{cedr} source code\footnote{https://github.com/Georgetown-IR-Lab/cedr} which contains the code implementation of CEDR-KNRM and BERT. For CONV-KNRM we use an adapted version of NN4IR code\footnote{https://github.com/faneshion/DRMM}. For ColBERT, we adopted the code released with the ColBERT paper\footnote{https://github.com/stanford-futuredata/ColBERT}.

\begin{itemize}[leftmargin=*]
\item
\textbf{CONV-KNRM}. The word embedding vectors are pre-trained and are fixed in CONV-KNRM. We use 300 dimension word embedding
vectors trained on TREC Disks 4 \& 5 or ClueWeb09-Cat-B
with skip-gram and negative sampling model~\cite{skipgram}. 
\item
\textbf{BERT}. 
The pre-trained BERT-base uncased model is  fine-tuned with an input as $[CLS] Q [SEP] D$. On top of the [CLS] vector representation, 
there is one feed-forward layer to combine the 768 dimensions to one score. 
\item
\textbf{CEDR-KNRM}. 
The CEDR-KNRM model is fine-tuned based on the trained BERT ranker, the number of kernels is 11 following the 
original paper~\cite{cedr}. 
\item
\textbf{ColBERT}. The ColBERT model is fine-tuned from the same initial weights as BECR as discussed before. 
The output dimension of linear projection is 128 as selected by the original paper. 
\end{itemize}

\subsection{Relevance and Inference Efficiency}
\label{sec:inferece:effect}

\begin{table*}[htbp]
	\caption{Relevance comparison of BECR against baselines. Significant improvements to $^{\ddag}$ConvKNRM, $^*$BERT, $^{\S}$ColBERT, $^{\P}$CEDR-KNRM at 95\% confidence level.}
	\centering
	\begin{small}
		\resizebox{2.1\columnwidth}{!}{
		\begin{tabular}{r || l | l |l  || l | l| l || l |l | l}
			\toprule
			Model & \multicolumn{3}{|c||}{ClueWeb09-Cat-B} 		& \multicolumn{3}{|c||}{Robust04} & MSMARCO & DL19 & DL20			\\
			  & NDCG@5  & NDCG@20 & P@20 & NDCG@5 & NDCG@20 &P@20 & MRR@10 Dev & NDCG@10 & NDCG@10\\
 			\midrule
			BM25	 & 0.2351 &  0.2294 & 0.3310 & 0.4594 & 0.4151 & 0.3548 & 0.167 & 0.488 & 0.480\\
			ColBERT (Ours) &  0.2408 & 0.2400 & 0.2067 & 0.3809  & 0.3498 & 0.3074 & 0.355 & 0.701 & 0.674\\
			ColBERT (from \cite{colbert,TRMD}) &  0.2273~\cite{TRMD} & 0.2365~\cite{TRMD} & 0.2507~\cite{TRMD}  & 0.4031~\cite{TRMD} & 0.3754~\cite{TRMD} & 0.3254~\cite{TRMD} & 0.349~\cite{colbert} & -- & -- \\
			CONV-KNRM & 0.2869$^{\S}$ & 0.2735$^{\S}$ &  0.3698$^{\S}$ & 0.4742$^{\S}$ & 0.4501$^{\S}$ & 0.3349$^{\S}$ & -- & -- & --\\
			BERT-base  & 0.2853$^{\S}$&  0.2612$^{\S}$ & 0.3764$^{\S}$ & 0.5160$^{\ddag\S}$ & 0.4514$^{\S}$ & 0.3983$^{\S}$ & 0.349 &  0.686 & 0.672 \\
			CEDR-KNRM (Ours) & 0.3030$^{\ddag\S}$ &  0.2693$^{\S}$ & 0.3961$^{\S}$ & 0.5563$^{*\ddag\S}$ & 0.4637$^{\S}$ & 0.4249$^{\S}$ & 0.344 & 0.702 & 0.686\\
			CEDR-KNRM (from \cite{cedr,2021Boystsov}) & -- &  -- & -- & -- & 0.5381 ~\cite{cedr}  & 0.4667~\cite{cedr}  & -- &
 0.682~\cite{2021Boystsov} & 0.675~\cite{2021Boystsov} \\
			\midrule
			BECR$^-$ & 0.3588$^{\P*\ddag \S}$ &  0.3066$^{\P*\ddag \S}$ & 0.4016$^{\S}$ & 0.5366$^{*\ddag\S}$ & 0.4635$^{\S}$ & 0.4045$^{\S}$ & 0.323 & 0.682 & 0.655\\
		BECR & 0.3632$^{\P*\ddag \S}$ &  0.3075$^{\P*\ddag \S}$ & 0.3987$^{\S}$ &  0.5349$^{\ddag\S}$ &   0.4656$^{\S}$ & 0.4005$^{\S}$ & 0.319 & 0.658 & 0.647 \\
			\bottomrule
		\end{tabular}
		\label{tab:main}
		}
	\end{small}
\end{table*}



\textbf{RQ1.} Table~\ref{tab:main} shows the relevance results  of our approach and the aforementioned baselines. 
Some table entries marked with a citation use relevance numbers published in the previous work as a reference point.
The LSH approximation is not applied in this table and we will discuss the impact of using LSH on BECR's relevance in Table~\ref{tab:storage}.
The BECR$^-$ model corresponds to the BECR model without using pre-computed token embeddings. 
Its representation of each query term is computed at runtime using the offline-refined BERT model. Thus BECR$^-$ combines the idea of ColBERT with BECR. The result of BECR$^-$ listed in the table shows that the relevance difference  between BECR and BECR$^-$ 
is relatively small within +/-1.2\% in all cases except that  there is a 3.6\% degradation for DL19.  We study this issue more in the next subsection.


Following previous work~\cite{cedr}, for Robust04 and ClueWeb09-Cat-B, we re-rank the 150 documents per query retrieved by Indri initial
ranking in the Lemur system\footnote{http://www.lemurproject.org/}. For the MS MARCO Dev set, we follow a common practice to
re-rank top 1000 passages per query retrieved by BM25.
For DL19 and DL20, their test query sets release top 100 passages per query and thus we re-rank these top 100 results. 

We discuss the comparison of BECR with the other baselines below in each dataset. 
The models with statistically significant improvements over the four baselines CONV-KNRM, BERT, ColBERT, and CEDR-KNRM, 
are marked with $^{\ddag}$, $^*$, $^\S$, and $^\P$, respectively. 
For example, NDCG@5 value 0.3632$^{\P}$  of BECR means it has a significant improvement over CEDR-KNRM at 95\% confidence level.

{\bf ClueWeb09-Cat-B and Robust04.}
On both datasets, 
the CEDR-KNRM model that combines BERT with KNRM significantly outperforms BERT and CONV-KNRM. 
That is consistent with the findings in \cite{cedr}.  
While BERT and CONV-KNRM also perform decently well, 
ColBERT doesn't perform as good as the other models. Our numbers for  ColBERT are consistent  with another
recent study~\cite{TRMD} that also trained ColBERT on Robust04 and ClueWeb09-Cat-B. 

Compared to the best performing baseline CEDR-KNRM, 
BECR outperforms it by a visible margin on ClueWeb09-Cat-B in NDCG@5 by 20\%  and NDCG@20 by 14\%, while it is slightly worse than 
CEDR-KNRM on Robust04 in NDCG@5 and P@20. 


{\bf MS MARCO.}
The characteristic of MS MARCO passage is distinctly different from ClueWeb09-Cat-B and Robust04 in which
it has longer queries and  much shorter document lengths.
The last 3 columns of Table~\ref{tab:main} show the performance for
MS MARCO passage ranking on its Dev set and TREC deep learning datasets. 
The result for the Dev set is for top 1000 passage re-ranking, and with top 150 re-ranking after initial BM25 ordering, MRR@10 for BECR is 0.301.
There is a visible relevance degradation  for BECR and the BECR$^-$ model compared to BERT and CEDR-KNRM.
We have sampled ranking results for a number of MS MARCO queries and find that because MS MARCRO has about 1 judgement 
label per query, which gives an inaccurate assessment when there  are multiple relevant answers to a query. 
TREC DL  sets with more judgement labels for MS MARCO passages capture more accurate performance difference and
relevance degradation is 6.7\% for DL19 and 6\% for DL20
comparing BECR with CEDR-KNRM in NDCG@10.


Overall speaking, based on the above discussions  on Table~\ref{tab:main}  for  all datasets and the results on the response time discussed below, 
BECR brings a reasonable tradeoff between  relevance and efficiency.
\textbf{RQ2.}
Table~\ref{tab:time} lists the average inference time in milliseconds on a GPU server or a CPU server when re-ranking top 150 matched documents, 
where the document length is 857 (average length of ClueWeb09-Cat-B corpus) and the query length $n$ is 3 or 5. 
We choose these query lengths to report because 
it is known that  the average query length in user queries of popular search engines is between two and three, and 
most of queries contain 5 or less words.
On the GPU server, we evaluate the models written in Python 3.6 with PyTorch 1.7.1. 
On the CPU server, we evaluate the models written in C++. 
For CEDR-KNRM, ColBERT, BECR$^-$ and BERT,  we only 
evaluate the Python/PyTorch code performance on GPU since they all use BERT during inference.
We include KNRM as an interesting reference point to demonstrate BECR's comparable inference efficiency to a fast pre-BERT neural ad-hoc model.
Notice the reported KNRM time can be further improved by LSH approximation~\cite{JiWWW2019}.

\begin{table}[htbp]
	\caption{Inference time and operation counts (FLOPs) to re-rank top 150 documents with average length 857 and  query length $n$ for 
ClueWeb09-Cat-B.}
	\centering
		\resizebox{1.02\columnwidth}{!}{
		\begin{small}
		\begin{tabular}{r || r | r | r | r }
			\toprule
			Model Specs. & n & FLOPs (ratio) & \multicolumn{2}{|c}{Time (ms) (ratio)} \\
			&&  & GPU & CPU \\
 			\midrule
 			KNRM & 3 & 148M ($5\times$) & 1.3 ($1\times$) & 123.5 ($5\times$)\\
 			& 5 & 246M($5\times$)  & 1.6($0.5\times$) & 312.8 ($8\times$)  \\
 			\midrule
 			ColBERT & 3 & 480M ($15\times$)  & 13.7 ($9\times$) & -- \\
 			 & 5 & 779M ($15\times$) & 13.7 ($4\times$) & -- \\
 			 \midrule
 			BERT & 3 & 12.2T ($234k\times$) & 4359 ($2900\times$) & --\\
 			 & 5 & 12.2T ($580k\times$) & 4431 (1300 $\times$) & --\\
 			 \midrule
 			CEDR-KNRM & 3 & 12.2T($234k\times$) & 5577 ($3700 \times$) & --\\
 			 & 5 & 12.2T ($580k\times$)& 5601 ( $1700 \times$) & --\\
 			\midrule
 			BECR$-$, L=13, LSH & 3 & 337M($11\times$)  & 6.3($4\times$) & --\\
 			& 5& 562M ($11\times$)& 10.5 ($3\times$)& -- \\
 			BECR$-$, L=5, LSH &  3 & 287M ($9\times$)& 3.0($2\times$)&  -- \\
 			& 5 & 478M ($9\times$)& 5.0 ($2\times$) & --\\
 			\midrule
 			BECR,L=13,LSH & 3 & 81M ($2.6\times$) & 2.9 ($2\times$) & 65.3 ($3\times$) \\
 			& 5 & 136M ($2.6\times$) & 5.7 ($2\times$) & 117.7 ($3\times$) \\
 			 BECR,L=5,LSH & 3 & 31M ($1\times$)&  1.5 ($1\times$) & 25.4 ($1\times$)\\
 			 & 5 & 52M ($1\times$) & 3.3 ($1\times$) & 40.7 ($1\times$) \\
			\bottomrule
		\end{tabular}
		\label{tab:time}
		\end{small}
		}
\end{table}

Table~\ref{tab:time}  
only reports the LSH-based BECR model and we will demonstrate its competitiveness in terms of  relevance in Table~\ref{tab:storage}.
The reported inference time only includes the score computation for the given number of top results to be re-ranked. 
The  I/O time for accessing embedding vectors discussed in Section~\ref{sect:discussonline} is not included in Table~\ref{tab:time}  
and it costs about 26ms and 32ms for re-ranking 150 ClueWeb09-Cat-B documents for BECR when L=5 and 13, respectively. 


In the presence of GPU, the models (BECR, ColBERT, BECR$^-$ and  KNRM) that don't involve BERT computation for 
documents only take a few milliseconds.  BERT and  CEDR-KNRM take a few GPU seconds which is too slow and expensive for interactive response.
Without GPU, the LSH version of BECR with 5 layers only takes tens of milliseconds while it takes  
up-to 117.7ms  with 13 layers  for queries with 5 tokens.
Thus the CPU-only execution of BECR with LSH is fast enough to be adopted in a search system for delivering a sub-second response time.

To further understand the computational cost, we trace the number of floating point operations 
(FLOPs) based on a PyTorch profiling tool called the torchprofile 
library\footnote{https://github.com/mit-han-lab/torchprofile}, and the third column of Table~\ref{tab:time} lists the 
operation counts reported by the tool.
The number reported by the tool might not reflect accurate operation counting. But still the traced numbers give us a good sense on the order of computation operations performed.
In particular, compared with the ColBERT model that also performs efficiently, The 5-layer BECR with LSH is $4\times$ or $9\times$
faster in inference time. The operation count reduction  ($15\times$ smaller in FLOPs) gives a supporting evidence.

\begin{figure}[h]

\includegraphics[width=0.8\columnwidth]{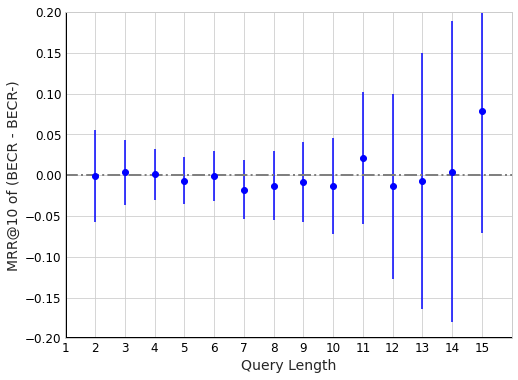}

\caption{MRR@10 difference between BECR and BECR$^-$, and 95\% confidence interval by query length  on MS MARCO Dev} 

\label{fig:len}
\end{figure}

\subsection{ Different Design Options for BECR}
\label{sec:eval:design}

\textbf{RQ3}.
\textbf{Impact of query decomposition and embedding composition by query lengths.}
Using pre-computed  token embeddings for a query
would reduce the contextual information that the ranker learns from the query,
and  we investigate to what extent it affects ranking performance based on different query lengths. 
We use the MS MARCO Dev set to compare BECR with BECR$^-$  because this set has a  large number of 
queries with diverse lengths.
Figure~\ref{fig:len} shows the MRR metric  difference between BECR and BECR$^-$ 
segmented by the query length, namely $MRR_{BECR} - MRR_{BECR^-}$.  
The mean value with a dot and the 95\%  confidence interval are  depicted for each query length.
For queries with 6 words or less, the BECR model performs similarly  
as BECR$^-$, with a 0.0015 drop over 4601 queries in MRR@10, while for the 2379 longer queries from 7 words to 15 words,
the average degradation of BECR from BECR$^-$ is 0.0069.

While BECR$^-$ seems to be slightly better than  BECR especially for longer queries in relevance, 
we are not able to confirm this conclusion with statistically significance for  this dataset when the query length increases
because the 95\% confidence interval at each query length overlaps with value 0. 
Overall, the query decomposition on longer queries has a bigger degradation than it on shorter queries but the degradation is of reasonable magnitude.


\begin{table}[htbp]
	\caption{Relevance of different design options for BECR with different query lengths measured by NDCG@5.} 
	\centering
	\label{tab:leaveout}
	\resizebox{1.02\columnwidth}{!}{
		\begin{small}
		\begin{tabular}{r || c | c | c | c | c }
			\toprule
			Model 	& \multicolumn{5}{|c}{ClueWeb09-Cat-B}	\\
			Query Length	& 1 & 2 & 3 & $\ge 4$ & Overall \\
 			\midrule
 			BECR; no token pairs & 0.3399& 0.4386 & 0.3228 &		0.3349 & 0.3537 \\
 			BECR; no uni-gram & -- & 0.4509 & 0.3235 &		0.3638 &	0.3617 \\
 			BECR; window=1 & 0.3407 & 0.4475 & 0.3108 &	0.3499 & 0.3520  \\
BECR; tokens for Q\&D & 0.298 &	0.4104 &	0.2543 &	0.2818 &	0.3034   \\
 			\midrule
 			BECR & 0.3407 & 0.4475 & 0.3250 & 0.3717 & 0.3628  \\
			\bottomrule
		\end{tabular}
		\end{small}
		}
\end{table}
\begin{table}[htbp]
	\caption{NDCG results after  removing a scoring component. Statistical significant degradation from BECR at confidence level 95\% is marked with $^{\$}$.}
	\centering
	\label{tab:component}
	\resizebox{1.02\columnwidth}{!}{
		\begin{small}
		\begin{tabular}{r || c | c | c }
			\toprule
			Model 	& \multicolumn{3}{c}{ClueWeb09-Cat-B} \\
			NDCG@k	& 3 ($\Delta \times 100$) & 5 ($\Delta \times 100$) & 10 ($\Delta \times 100$) \\
 			\midrule
 			BECR & 0.3754 & 0.3632 & 0.3417 \\
 		    \midrule
 			No $S_{deep}$ & 0.2915 (-8.4)$^{\$}$ & 0.2948 (-6.8)$^{\$}$ & 0.2788 (-6.3)$^{\$}$ \\
 			No $S_{lexi}$ & 0.2673 (-10.8)$^{\$}$ & 0.2643 (-9.9)$^{\$}$ & 0.2561 (-8.6)$^{\$}$ \\
 			No $S_{others}$ & 0.3099 (-6.5)$^{\$}$ & 0.3032 (-6.0)$^{\$}$ & 0.2922 (-5.0)$^{\$}$ \\
			\bottomrule
		\end{tabular}
			\end{small}
		}
	
\end{table}

\textbf{Other design options.}
Table~\ref{tab:leaveout} compares other  design options for BECR  when the number of query words varies in ClueWeb09-Cat-B.
The row marked with ``BECR; no token pairs'' means online query representation is derived only based on uni-grams. 
This is the extreme case where the weights on all token pairs are 0. BECR outperforms this variation visually for the overall NDCG@5 score, and especially for longer queries in this case. 
This shows that the use of token pair encoding to approximate query representation brings a visible advantage.
On the other hand, this result shows that when the embeddings of word-pair tokens are not available,
BECR can resort to rely on uni-gram embeddings to compose the query representation. 

The row marked with ``BECR; no uni-gram'' means online query representation is derived only based on token pairs, and 
the  weights on uni-grams are 0. BECR has a comparable relevance in this case, meaning the use of 
uni-gram embeddings does not bring a significantly advantage. 
This is intuitive as the word-pair embeddings contains the meaning of uni-grams. However as stated above, the use of uni-grams is still  meaningful especially when the 
embeddings for some word pairs are not pre-computed.

The row marked with ``BECR; window=1'' means we only consider uni-gram and adjacent query
word pairs in the encoding. By comparing the result of it with $BECR$ on longer queries (column $3$ or $\ge 4$), we can see that incorporating the nonadjacent word pairs seem to help improve the quality of query representations.


Lastly, the row marked with ``BECR; tokens for Q\&D'' represents a case 
where both the query and the document are encoded using pre-computed uni-gram and word-pair embeddings during inference time. 
The model relevance drops significantly, indicating that document embeddings  directly composed from pre-computed token embeddings
are less effective.

\textbf{RQ4.} 
Table \ref{tab:component} reports the NDCG results after removing one of scoring components to demonstrate the contribution of each component in ClueWeb09-Cat-B. 
Taking BECR as baseline, for each model variation after removing a component, we report the NDCG results with  
the NDCG degradation $\Delta$ multiplied by 100.
Statistical significant degradation from BECR at confidence level 95\% is marked with $^{\$}$.
For example, for Row marked with ``No $S_{deep}$'', it means the model trained without the deep soft contextual matching component and
there is a 6.3 point drop in NDCG@10.  The results for the other datasets have a similar pattern.

From Row``No $S_{lexi}$'', the removal of $S_{lexi}$ causes 
relevance  degradation by 10.8, 9.9, and 8.6 points for NDCG@3, NDCG@5, and NDCG@10, respectively in ClueWeb09-Cat-B. 
For the other datasets,  we observe a 3\% degradation on Robust04 and  a 5\%-7\% drop on the MS MARCO Dev and TREC DL datasets.
Thus, adding the lexical component into BECR brings benefits for both document and passage re-ranking. 
BECR relies more on $S_{lexi}$ for ClueWeb09-Cat-B, probably because
the content of web data is more diversified with mixed qualities, and there are more matched documents for each query. 
\comments{
The previous work typically  depends on the deep neural network  to fully catch both lexical and semantical relatedness,  
the above result shows that both  classical ranking signals and  deep neural signals play a significant role in BECR. 
}

The removal of $S_{others}$ causes a noticable  degradation of NDCG results for ClueWeb09-Cat-B. That is because $S_{others}$ contains the pagerank score, 
which is a very useful document quality signal. 

\textbf{RQ5.}
We evaluate the tradeoff between storage reduction and relevance with LSH and compact layer choices of BERT.
Table~\ref{tab:storage} shows the NDCG relevance score and storage space requirement for different LSH encoding size and 
the number of layers used for ClueWeb09-Cat-B which has more and longer documents than the other tested datasets.
\comments{
First, we fix the number of layers to be 13. From the table we can see that a 256-bit encoder will have comparable relevance score in NDCG@3, NDCG5, and NDCG@10 compared to the original model. This is a great news because we will be able to reduce the storage size by almost $100\times$ without visible relevance impact. 

Next we reduce the number of layers used in supplying the embedding vectors for  the online inference. Note that when $L=5$, we use the embedding layer, the first 3 encoder layers and the last encoder layer. When $L=1$ corresponds to only use the last encoder layer. We separately train a BECR model for $L=13,5,1$, and we report the NDCG@k on the test set with the original BERT representation and the representation encoded into a $b$-bit LSH footprint.

When choosing LSH with 256 bits,
there is a small relevance degradation from 13 layers to 5 layers for NDCG@3 and NDCG@10 and the difference in NDCG@5  is less visible. None of the degradation is statistically significant. 
Thus it is very feasible to reduce the number of representations from 13 to 5, which result in an additional 2.6$\times$ 
reduction in storage space if needed. 
}
From the table we can see that by LSH encoding with 256-bit footprint and using 5 layers in training and inference, we are able to reduce the document storage to $0.4\%$ of the original space while achieving comparable ranking performance.


\comments{
\begin{table}[htbp]
	\caption{Relevance (drop in percentage) and storage tradeoff (ClueWeb09-Cat-B).  Significant drop from original L=13 at confidence level 95\% is marked with $^{\$}$.}
	\centering
		\begin{small}
		\begin{tabular}{c  c || c | c | c || c | c }
			\toprule
			LSH bit	&  & \multicolumn{3}{|c||}{NDCG@k} 	& \multicolumn{2}{|c}{Storage} 			\\
			b & L & 3 & 5 & 10 & D(TB) & Q(TB)\\
 			\midrule
 			Original & 13 & 0.3706 & 0.3537 & 0.3334 & 1711 & 37.9 \\
 			512 &  13 & 0.3613 & 0.3506 & 0.3284 & 35.8 & 0.79\\
 			256 &  13 & \textbf{0.3723}& \textbf{0.3550} & \textbf{0.3341}  & 17 & 0.39\\
 			128 &  13 & 0.3376$^{\$}$ & 0.3319 & 0.3205 & 9.1 & 0.20\\
 			64 &  13 & 0.3256$^{\$}$ & 0.3191$^{\$}$ & 0.3143 & 4.6 & 0.10 \\
 			32 &  13 & 0.3065$^{\$}$ & 0.3057$^{\$}$ &0.2952$^{\$}$ & 2.4 & 0.05\\
 			\midrule
 			Original & 5 & 0.3672  & 0.3595 & 0.3375 & 658.1 & 14.6 \\
 			\textbf{256} & \textbf{5} & \textbf{0.3572} & \textbf{0.3546} & \textbf{0.3268} & \textbf{7.0} & \textbf{0.15}\\
            Original & 1 & 0.3348$^{\$}$ & 0.3263$^{\$}$ & 0.3112$^{\$}$ & 76.9 & 1.36 \\
            256 & 1 & 0.3338$^{\$}$ & 0.3262$^{\$}$ & 0.3111$^{\$}$ & 1.5 & 0.03 \\
			\bottomrule
		\end{tabular}
		\label{tab:storage}
		\end{small}
\end{table}
}

\begin{table}[htbp]
	\caption{Relevance (drop in percentage) and storage tradeoff (ClueWeb09-Cat-B).  Significant drop from original L=13 at confidence level 95\% is marked with $^{\$}$.}
	\centering
		\begin{small}
		\begin{tabular}{c  c || c | c | c || c | c }
			\toprule
			LSH bit	&  & \multicolumn{3}{|c||}{NDCG@k} 	& \multicolumn{2}{|c}{Storage} 			\\
			b & L & 3 & 5 & 10 & D(TB) & Q(TB)\\
 			\midrule
 			Original & 13 & 0.3754 & 0.3632 & 0.3417 & 1711 & 37.9 \\
 			512 &  13 & -2.5\% & -0.9\% & -1.5\% & 35.8 & 0.79\\
 			256 &  13 & \textbf{+0.4\%}& \textbf{-0.9\%} & \textbf{+0.2\%}  & 17 & 0.39\\
 			128 &  13 & -8.9\%$^{\$}$ & -6.2\% & -3.9\% & 9.1 & 0.20\\
 			64 &  13 & -12\%$^{\$}$ & -9.8\%$^{\$}$ & -5.7\% & 4.6 & 0.10 \\
 			32 &  13 & -17.3\%$^{\$}$ & -13.6\%$^{\$}$ &-11.5\% $^{\$}$ & 2.4 & 0.05\\
 			\midrule
 			Original & 5 & 0.3672  & 0.3595 & 0.3375 & 658.1 & 14.6 \\
 			\textbf{256} & \textbf{5} & \textbf{-2.7\%} & \textbf{-1.4\%} & \textbf{-3.2\%} & \textbf{7.0} & \textbf{0.15}\\
 			\midrule
            Original & 1 & 0.3348$^{\$}$ & 0.3263$^{\$}$ & 0.3112$^{\$}$ & 76.9 & 1.36 \\
            256 & 1 & -0.3\% $^{\$}$ & -0.0\% $^{\$}$ & -0.0\% $^{\$}$ & 1.5 & 0.03 \\
			\bottomrule
		\end{tabular}
		\label{tab:storage}
		\end{small}
\end{table}

\section{Concluding Remarks}
\label{sec:conclusion}

The main contribution of this work is  to  show  how a lightweight BERT based re-ranking model for top results with 
decent  relevance can be made possible
to deliver fast response time on affordable computing platforms.  
Our evaluation shows that its inference time using C++ on a CPU server without GPU is as fast as tens of milliseconds for the ClueWeb09 dataset 
while the access to an embedding store incurs a modest time cost. 
With composite token encoding, 
BECR effectively approximates the query representations and makes a use of both deep contextual and  lexical matching features, 
allowing for a reasonable tradeoff between  ad-hoc ranking relevance and efficiency.  

There are  opportunities for additonal improvement as future work. For example,
the disk access cost to fetch precomputed embeddings still takes a significant amount of time compared to the  reduced
ranking inference time, and  when there is a large number of queries processed simultaneously, disk I/O contention can arise.
Thus further embedding storage optimization is desirable.

{\bf Acknowledgments}. We thank Mayuresh Anand, Shiyu Ji, Yang Yang,  and anonymous referees
for their valuable comments and/or help. 
This work is supported in part by NSF IIS-2040146  and by a Google faculty research award.
It has used the Extreme Science and Engineering Discovery Environment~\cite{XSEDE}
supported by NSF ACI-1548562.
Any opinions, findings, conclusions or recommendations expressed in this material
are those of the authors and do not necessarily reflect the views of the NSF.

\newpage

\bibliographystyle{ACM-Reference-Format}
\bibliography{bib/ref}
\newpage

\appendix

\section{Additional evaluation information }
\label{sec:data}

\noindent
\textbf{Data}. The data collections used are summarized in Table~\ref{tab:data}.
1) Robust04 uses TREC Disks 4 \& 5\footnote{https://trec.nist.gov/data/robust/04.guidelines.html} (excluding Congressional Records). Note that we used
title queries for the reported experiments.
To train the model, we split the queries into five folds based on \cite{Huston2014ACO}, with three folds used for training, one for validation, and one for testing.
2) ClueWeb09-Cat-B uses ClueWeb09 Category B with 50M web pages. There are 200 topic queries from the TREC Web Tracks 2009 to 2012\footnote{https://lemurproject.org/clueweb09/}. After removing one query with 10 terms, the lengths of queries range from 1 to 5.
Spam filtering is applied on ClueWeb09 Category B using Waterloo spam 
score~\footnote{from the fusion set in https://plg.uwaterloo.ca/~gvcormac/clueweb09spam/} with threshold 60. 
To ensure enough samples in training phase, we add the 50 queries from 2012 in all the five training folds.
For the 150 queries from 2009 to 2011,  we split the queries into five folds based on \cite{Huston2014ACO}.
3) The MS MARCO passage ranking dataset\footnote{https://github.com/microsoft/MSMARCO-Passage-Ranking}  collects the passages from Bing search logs, consisting of 8.8 million passages and over 50 thousand queries with relevance labels. Our evaluation also uses
TREC DL 2019 and 2020\footnote{https://microsoft.github.io/msmarco/TREC-Deep-Learning-2020.html} from MS MARCO as they have more
comprehensive relevance judgements.

\begin{table}[htbp]
    \caption{Summary of Evaluation Data}
    \centering
    \resizebox{1.02\columnwidth}{!}{
    \begin{tabular}{|c|c|c|c|c|c|}
        \hline
                Dataset & Domain & \# Query & \# Doc & Query Length & Mean Doc Length \\
        \hline
        ClueWeb09 & Web & 149 & 50M & 1-5 & 857 \\
        Robust04 & News & 250 & 0.5M & 1-4 & 479 \\
        MS MARCO Dev & Q\&A, passage & 6980 & 8.8M & 2-15 & 57\\
        TREC DL19 & Q\&A, passage & 43 & -- & 2-15 & -- \\
        TREC DL20 & Q\&A, passage & 54 & -- & 2-15 & --\\
        \hline
    \end{tabular}
    }
    \label{tab:data}
\end{table}

\noindent
{\bf Non-neural features}.
Table \ref{tab:leaveout} lists the non-neural features used in our three-dataset evaluation.  Feature
``tfidf in X'' means the TF\-IDF score of each query token in X where X is either body or title.  Feature
``bm25 in X'' means the BM25 score of each query 
token in X.  Feature ``inv\_min\_dist in X'' means the inverse of the smallest distance between any pair of query tokens in X. 
The prefix max, min, avg and sum mean the maximum, minimum, average or sum of the corresponding feature values for all query tokens.

\begin{table}[htbp]
	\caption{Non-neural features used in the evaluation} 
	\centering
	\label{tab:leaveout}
	\resizebox{0.9\columnwidth}{!}{
		\begin{tabular}{l | l }
			\toprule
            Feature name & Note\\
 			\midrule
            max tfidf in title & \multirow{4}{*}{not in MS MARCO passage}\\
            min tfidf in title & \\
            avg tfidf in title & \\
            sum tfidf in title & \\
            \midrule
            max tfidf in body & \\
            min tfidf in body & \\
            avg tfidf in body & \\
            sum tfidf in body & \\
            \midrule
            sum tfidf in body and title & not in MS MARCO passage \\
            \midrule
            max bm25 in title & \multirow{4}{*}{not in MS MARCO passage}\\
            min bm25 in title & \\
            avg bm25 in title & \\
            sum bm25 in title & \\
            \midrule
            max bm25 in body & \\
            min bm25 in body & \\
            avg bm25 in body & \\
            sum bm25 in body & \\
            \midrule
            sum bm25 in body and title & not in MS MARCO passage\\
            \midrule
            max inv\_min\_dist in title & \multirow{3}{*}{not in MS MARCO passage}\\
            min inv\_min\_dist in title & \\
            avg inv\_min\_dist in title & \\
            \midrule
            max inv\_min\_dist in body & \\
            min inv\_min\_dist in body & \\
            avg inv\_min\_dist in body & \\
            \midrule
            pageRank & only available in Clueweb09\\
			\bottomrule
		\end{tabular}
		}

\end{table}

\section{An example of ranking with BECR}
\label{sec:expl}
\begin{figure*}[htbp]
    \centering
   \includegraphics[width=\textwidth]{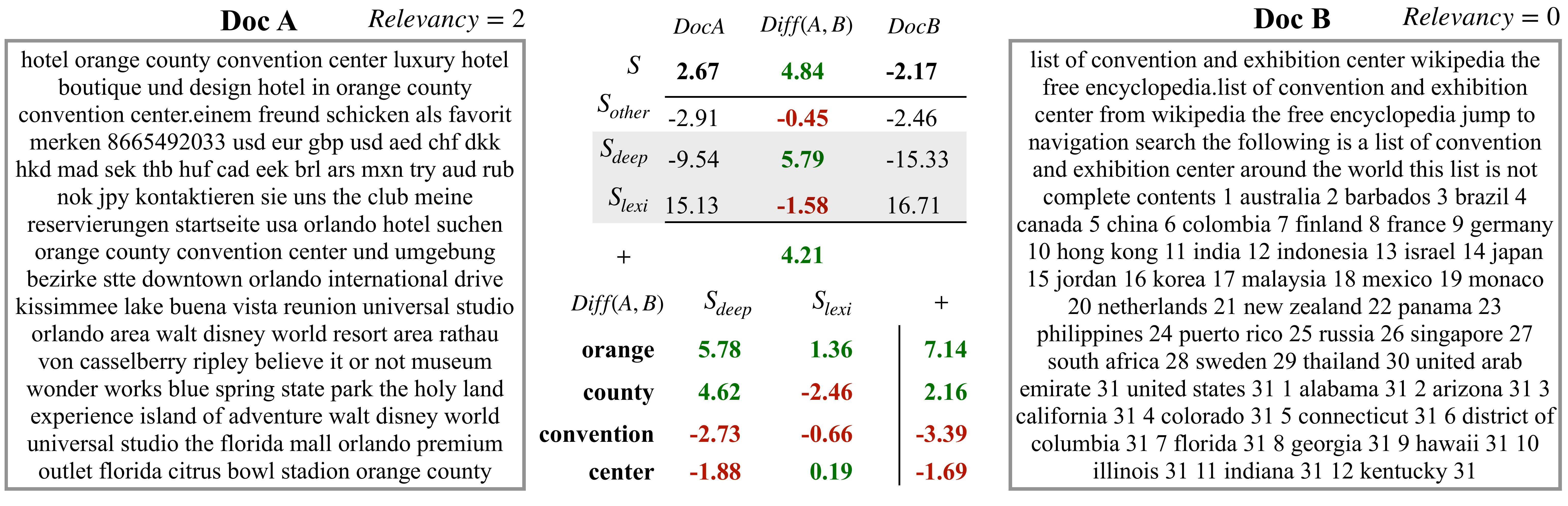}
    \caption{Score difference and breakdown of two matched documents for Query "orange county convention center"}
    \label{fig:ex_good}
\end{figure*}

This section gives an example to rank two documents for a query with BECR and  illustrates its component scores in
Figure~\ref{fig:ex_good}.
Taking advantage of linear composition, we use the score and subscore difference to 
understand feature contribution  in  comparing the document rank order. 
Both the query and documents are from the ClueWeb09-Cat-B dataset. The query is 
\textit{orange county convention center}. Document  $A$ is judged as  relevant while  Document $B$ is not, and BECR gives a higher final rank 
score to Document A.
Based on the text of documents, 
Document $A$ is about hotel information near Orange county convention center while 
Document $B$ is a Wikipedia article with a list of convention and exhibit centers. 

The middle of  Figure~\ref{fig:ex_good} depicts the score differences of these two documents. 
\comments{
As we can see, the ranking score is the sum of the four component scores. Based on the difference between the ranking scores, 
we can easily tell that although the Doc B has a high document score than Doc A (the difference is 0.45). By pulling together 
the $S_{deep}$ and $S_{lexi}$ (the $+$ row), we can tell that Doc A matches with the query better. 
The preferences jointly lead to a final ranking preference of Doc A over Doc B. 
}
The overall score of Document A as 2.67 is higher than that of Document B. The dominating reason is that $S_{deep}$ deep soft matching score of $A$ is much higher than that of $B$ even the lexical component score  of  $A$ is slightly lower than $B$.
That can be observed by looking at the largest difference in the $Diff(A,B)$ column.
 
To further understand why Document $A$ matches better semantically for this query, 
we further show the subscore breakdown of soft and lexical matchings for the token interaction level in the bottom table.
The column under $S_{deep}$ shows the difference breakdown $Diff(A,B)$  for component $S_{deep}$ under each query token word. 
while the column under $S_{lexi}$ shows the difference breakdown for $S_{lexi}$. 
The last column under $+$ contains the sum of  the difference under each token by adding up the soft and lexical matching subscores. 
We can see that Document $A$ matches the query better because it has more information regarding \textit{orange county}, 
although Document  $B$  matches with \textit{convention center} better. 
Token \textit{orange} is not mentioned at all 
in Document $B$ and this is the reason  that both soft and lexical matching prefers Document $A$. 
For token \textit{county}, Document $B$ 
has a higher lexical matching subscore because that term occurs several times in Document $B$. 
\comments{
On the contrary, 
the deep matching score have a stronger preference for Doc A. We think it is probably due to the fact that the deep 
contextual token representations are able to capture the specially meaning in the co-occurrence of the term 
\textit{orange} and \textit{county} in the document. For the tokens \textit{convention center}, we would 
expect Doc B to have better score as the whole article is about convention centers. Indeed, the 
model prefers Doc B, although the difference for that two tokens are not as significant as \textit{orange country}. 
}

\comments{
The above example shows  a simplicity  of composition linearity in explaining the contribution of each feature to the overall ranking.
The PageRank score of each document as a feature. 
The weight of PageRank feature 
learned by the model is 0.044. 
So if a document's PageRank score increase by 1, we would expect the ranking score to increase by 0.044. 
}

The PageRank score is 4.69 for document $A$ and  
6.77 for Document $B$.  The weight of PageRank feature 
learned by BECR is 0.044. Based on linearity,  PageRank feature contribute 
0.92 ($0.044 * (6.77-4.69) = 0.92$) to the score difference (both $S_{others}$ and $S$) between Doc A and Doc B. It is not sufficient to outweigh the difference caused by soft matching.

\end{document}